# Line Emission in AGN: Effects of Local Delays Upon Line Variability

Jason A. Taylor[1]

Code 661
Laboratory for High Energy Astrophysics
NASA/Goddard Space Flight Center
Greenbelt, MD 20771, USA

## ABSTRACT

Previous works concerning active galactic nuclei (AGN) variability (e.g., Blandford & McKee 1982) have assumed that the emission characteristics of illuminated clouds are purely a function of the instant continuum flux to which they are exposed. This paper shows that this assumption is not necessarily justified and that the history of exposure accounting for "local delays" due to finite cloud equilibrium times can also be relevant. For this reason, a new formalism is developed in this paper for computing the observational properties of models which have local delays. The nature of the nonlinear behavior that results is calculated for some very simple nonlinear cloud line emission models. It is found that the mean response time is a function of the recent average value of the continuum. Linear models fit to these nonlinear systems respond too slowly when there are low-energy (and generally rapid) changes in the continuum, yet respond too rapidly when there are high-energy (and generally slow) changes in the continuum. As with systems without local delays, the expression for the time-dependent line flux contains an integration over history of the "spatial" response function, which has structure at lags of the light travel times of the emission region. However, the kernel of this integral itself is a function of additional integrations over individual "cloud" response functions which have structure at lags of the equilibrium times of the cloud properties relevant to line emission. In the linear regime, the response can be approximated using a single response function. The integral of this function over lag is not generally equal to the mean flux in the line. Rather, it differs by a factor that is the strength of response for low-frequency continuum

---

[1]Also Department of Physics, University of Maryland, College Park, MD 20742, USA

excitations or simply the "asymptotic gain," which is unity only in fully linear models. If instantaneous or linear response is incorrectly assumed, local delays and nonlinear response can make a system appear larger than it actually is. These effects are similar to those that beaming can cause. Local delays can also be a source of asymmetry about the peak of the cross-correlation function.

*Subject headings:* galaxies: Seyfert—quasars: absorption lines—quasars: emission lines—quasars: general

## 1. INTRODUCTION

This is the first of five papers about a detailed investigation of the viability of three well-defined active galactic nuclei (AGN) cloud models. In order to follow the models to their logical conclusions, development of a theory general enough to accommodate many of the physical processes to be included was required. One of these processes is the effect that finite cloud response times can have upon the line variability that is observed. This effect, which has not been considered in any previous works, is the primary focus of this paper.

Procedures for computing a linearized response function of the time-dependent line emission given off from an ensemble of clouds illuminated by a time-dependent source are well known (e.g., Blandford & McKee 1982). They assume that the contribution toward line emission from a specific source is purely a function of the radius from the central object and the immediate continuum flux it is subjected to. This requires that the processes relevant to its line emission attain equilibrium much more quickly than the other time scales involved. The explicit time-dependent response of individual clouds, where, e.g., the line emission efficiency in a cloud lags the continuum flux it experiences, has not yet been accounted for in previous works concerning AGN variability.

However, accounting for finite equilibrium times can yield interesting results for most of the AGN cloud models that have been proposed. Consider, for instance, a cloud model in which the cloud area is a decreasing function of the cloud pressure, which is externally regulated by the pressure of an inter-cloud medium. Rees, Netzer, and Ferland (1989) additionally assumed $P \propto r^{-s}$, where $P$ is the pressure throughout the cloud and $r$ is the distance from the black hole. Let us consider the analogous case where the pressure is regulated by the local ionizing continuum flux $F_c$ and only indirectly through $r$, namely $P \propto F_c^{s/2}$. Such a dependence implies that a change in the continuum luminosity invokes a change in the cloud pressure as well. As we shall see, "reactive" cloud models like this one offer both theoretical and empirical advantages over static ones. Note that the clouds would not react instantaneously; a minimum for the characteristic time scale for internal pressure equilibrium to be asymptotically obtained is the sound crossing time of the clouds. As noted in Netzer (1990), this time scale can be of order of the continuum variation time scales, suggesting that clouds of this model might rarely be in actual pressure equilibrium. Therefore, even though the outermost layer emitting a line can be a small fraction of the cloud as a whole, clouds of this model should to some extent "remember" their prior pressures and areas.

Because line emission from clouds is a strong function of the area, pressure, and pressure ionization parameter $\Xi$ (defined here as the ionizing photon to gas pressure ratio), the line efficiency of a cloud has a nontrivial time dependence. For instance, consider the case where the continuum flux local to a cloud suddenly increases. If $s < 2$, the pressure ionization parameter of the cloud would at first follow the increase in the continuum flux, but would then decrease as the pressure begins to approach its new equilibrium value. Relative to Ly$\alpha$, the flux in a line like N v $\lambda$1240, which is probably a relatively high ionization transition in stable cloud sections (Taylor 1994), would initially rise, but then decay as the ionization parameter decreases. The response function that one would obtain upon a linear fitting would have structure not only at the range of lags corresponding to the light crossing times of the emission region, but also at lags greater than these by the pressure equilibrium times in the clouds.

In such a case, the previous works on AGN variability, which have all assumed that a response function at a given lag is proportional to the density of clouds along the corresponding "iso-delay" surface, are inapplicable. Specifically, the results based upon equation (2.13) of Blandford & McKee (1982), which was derived under the assumption that the equilibrium time scales of the cloud properties are all much less than the light crossing time (hereafter, the "fast cloud" assumption), are now suspect. This is an important point because a great deal of effort has been expended to obtain and analyze variability data using the approach of Blandford & McKee (1982).

In § 2 of this paper we shall find that there are several cloud properties affecting line emission that could be strong functions of the local continuum flux with equilibrium times large enough to violate the fast cloud assumption. Because for these cases the popular formalism of Blandford & McKee (1982) is inapplicable, in § 3 a new and more general formalism for analyzing variability data will be developed. This new formalism is compatible with models that have clouds with finite equilibrium times and nonlinear response. It will also be used in the other papers in this series (Taylor $1996a - d$; hereafter, Papers II-V). Readers not interested in the mathematical derivation of the time-dependent line profile with the new formalism may wish to skip to § 4, where the new theory is applied to some simples models. A summary is provided in § 5.

In Paper II, methods of measuring the response characteristics of clouds assumed to be in orbital motions will be presented. In Paper III, predictions made from applying the theory to three well-defined AGN cloud models using the values for the time constants computed in § 2 will be compared to existing variability data. A similar analysis will be provided in Paper IV, but will be directed just toward the character of the emission and absorption line shifts. In Paper V, the main results of the study will be presented, where the three models will be quantitatively tested by fitting them to data of NGC 5548, with the intent of keeping the freedom in parameter space granted to each model as equal as possible.

## 2. MOTIVATION

For the formalism of Blandford & McKee (1982) to be *in*applicable for a given cloud model, two conditions must be satisfied for at least one of the cloud properties in the model. The first of these conditions is that the line emissivity be a moderately strong function of the cloud property and that the cloud property in turn be a moderately strong function of the local continuum flux a cloud experiences. The second condition is that the equilibrium time scale of the cloud property be near one of the other characteristic time scales of the system. If the equilibrium time scale is near or greater than the line emission region light crossing time, the response function will be affected. Furthermore, if the equilibrium time is near the time for clouds to cross the emission region, the time-averaged line profile can be affected. Determining the precise way in which the response functions and profiles are affected requires a detailed and highly model-dependent analysis. Before going through such an analysis, let us first discuss some of the cloud line emission model properties which apparently meet the above two conditions.

Table 1 lists some of the processes responsible for reactive cloud properties in several of the models that have been proposed and the equilibrium time scales associated with them. Also shown is whether the slowness of equilibrium affects the response functions, line profiles, or line ratios. The first entry is for a two-phase pressure-equilibrium model (e.g., Wolfe 1974; Krolik, McKee, & Tarter 1981). Assuming in this case that the cloud pressure is regulated by pressure of the inter-cloud medium, the delay in the cloud pressure response to the continuum is limited by the inter-cloud temperature equilibrium time scale. For the model parameters described in Table 1, this is (only) $\simeq 43$ days. If the dependence of the inter-cloud temperature upon the local continuum flux is strong enough, the responding pressure will affect the response functions for the parameters assumed in Table 1 in a highly line-dependent fashion, giving the line ratios a complicated time dependence. Furthermore, if the cloud identities are preserved (as in Rees, Netzer, & Ferland 1989), the slowness of the cloud area and column density reactions will also affect the response functions respectively in a line-independent and weakly line-dependent fashion. The time-averaged line profiles for this model are not affected by the finite pressure equilibrium time, which is too small compared to the cloud crossing time ($\sim 2$ years for the parameters shown in Table 1) to be affected. However, if the inter-cloud temperature dependence is moderately strong, this model, like several others that are not immune to the various processes analyzed in Table 1, requires use of a new formalism. Such a formalism will be developed in § 3.

Note that the physical processes considered in Table 1 were drawn from the set of processes invoked by the various cloud models that have been proposed. In principle, all of these could be incorrect. Therefore, Table 1 is necessarily incomplete. For this reason, the analysis of time-dependent cloud response could be important even if all of the processes in Table 1 somehow accommodated the fast cloud assumption. As will be shown in Paper IV, such analysis does in fact constrain the mechanisms that are permissible in AGN models.

## 3. THEORY OF RESPONSE

In this section we shall extend the formalism in Blandford & McKee (1982) so that information can be obtained from variability data about models which violate two fundamental assumptions made in Blandford & McKee (1982): (1) instantaneous response and (2) linear response.

### 3.1. Locally-Delayed Response

Before one can understand the overall, global response of systems that can violate the fast cloud assumption, one must first understand the response of the individual clouds that make up such systems. In this section a general method of determining the time dependence of an arbitrary cloud property is derived. In § 3.2, this method will be used to obtain the global response of systems that can violate the fast cloud assumption.

The character of an AGN cloud's response depends critically upon the relative magnitude of two time scales. One of these is the variation time scale of some condition externally imposed upon the cloud, such as the local continuum flux or inter-cloud pressure. Another is the characteristic equilibrium time scale of a physical property of the cloud, such as its temperature or size, in response to the variations of the external conditions. As an example, let us consider the case of a cloud with a physical property that is an increasing function of the local continuum flux. Let us also assume for this example that the continuum source itself is time-independent, but that the cloud is in a periodic orbit about the black hole. This situation is a simple variation of those in which the continuum does vary. If we assume that the orbital period is significantly greater than the property's equilibrium time scale, the physical property would lag the time-averaged continuum flux to which the cloud is exposed as it orbits the black hole. The line emission in such a cloud would depend not only on its position, but also on another variable which indicates its orbital phase. Under certain conditions (see Appendix A) it can be shown that this variable can be the cloud velocity vector. For instance, if the line emissivity of a cloud is an increasing function of just the physical property, then the line emission from the cloud would be greatest not when it is closest to the black hole, but slightly farther away, after the cloud has acquired a small outward velocity. As the orbital time becomes even larger, we approach the "fast cloud" regime. In this regime we can assume that the physical property in the cloud reacts fast enough that it is purely a function of the local flux or in this case distance from the black hole such that the phase lag is zero.

A second case to consider is one where the variation time scale of the local continuum flux at a cloud is significantly smaller than the property's equilibrium time. Here the physical property of the gas would lag the orbital motion by a significant phase. This implies that a line with a strong enough dependence upon the lagged property could attain maximum flux when the cloud has a relatively high outward radial velocity. In this case, as the variation time scale of the input continuum flux becomes even smaller, we approach the "slow cloud" regime, where we can simply assume that the relevant physical property is a constant throughout the orbit.

The third possible case to consider is one in which the local continuum variation time scale is intermediate and of order of the equilibrium time scale. Understanding this case requires a more quantitative approach than the other two cases. Let us call the generic cloud property of interest $y(t)$ where $t$ is the time measured in the reference frame of the observer. The analysis which follows is quite general and $y(t)$ could represent properties such as the mean cloud area, pressure, or column density. Similarly, let $x(t)$ be a generic input, such as the local continuum flux near the cloud, of which $y(t)$ is assumed to be a function. Let $y'(x)$ be the asymptotic functional dependence of $y$ upon $x$ once sufficient time has elapsed for equilibrium to be established, where the prime denotes the functional dependence in the fast cloud regime. Let us furthermore assume that there exists a characteristic time scale $\tau_y$ for $y$ to respond to changes in $x$. Such a characteristic time will be equal to the ratio of the extent to which $y$ is out of equilibrium to the rate at which the non-instantaneously responding component of $y$ actually attains its equilibrium value. This gives

$$\tau_y = \frac{y'(x(t)) - y(t)}{\dot{y}(t) - \dot{x}(t)\tilde{\tilde{\Psi}}_{y|x}(\infty) <y>/<x>}, \tag{3-1}$$

where $\tilde{\tilde{\Psi}}_{y|x}(\infty)$ is a free parameter that is the instantaneous component of the "gain" of $y$ with respect to $x$, and $<x>$ is the average or "bias" of $x$, etc. The gain itself is an operator (defined by eq. [B3]) that yields the dimensionless ratio of the amplitudes of small variations of an output about its mean with respect to that of some input. It is merely the Fourier transform of the linearized response function (see Appendix B). The implicit assumption here that $\tau_y(x)$ is approximately constant could be invalid under the following conditions: the variations of $x$ are large enough, the initial conditions are far enough from equilibrium, or the equilibrium time has an explicit dependence upon the sign of $\dot{x}(t)$. In these cases the physics associated with the response time is not properly described by only one parameter. Otherwise, equation (3-1) completely characterizes the system given the prior inputs $x(t' \leq t)$ and the other system characteristics $y'(x), \tau_y$, and $\tilde{\tilde{\Psi}}_{y|x}(\infty)$.

Though we will assume that $y'(x)$ is a nonlinear function, in certain instances we shall find it highly instructive to consider the case in which the variations in $x$ are small enough that $y'(x)$ is accurately described by a first-order Taylor expansion. Performing such linearization of equation (3-1) (with eqs. [B1]-[B4]) yields in the frequency domain

$$\hat{\tilde{\Psi}}_{y|x}(\omega) = \frac{\eta(y|x) + i\hat{\tilde{\Psi}}_{y|x}(\infty)\tau_y\omega}{1 + i\tau_y\omega}, \tag{3-2}$$

where $\omega_y \equiv 1/\tau_y$ and $\eta(y|x) \equiv \hat{\tilde{\Psi}}_{y|x}(0)$ is the "asymptotic gain" of $y$ with respect to $x$ (see also eq. [B2]). Equation (3-2) can be used to formulate a more precise definition of the fast and slow cloud regimes, which respectively occur for $|\omega| \ll \omega_y$ and $|\omega| \gg \omega_y$, where the transfer function becomes a trivial function of $\omega$ (flat in $\log - \log$ coordinates).

Equation (3-2) yields in the time domain

$$\hat{\Psi}_{y|x}(\tau) = \hat{\tilde{\Psi}}_{y|x}(\infty)\delta(\tau) + \Theta(\tau)[\eta(y|x) - \hat{\tilde{\Psi}}_{y|x}(\infty)]\omega_y e^{-\tau\omega_y}, \tag{3-3}$$

where $\Theta$ is the step function. This response function tells us (namely via eq. [B6]) the contribution in the linear regime toward the output (e.g., the cloud area) made by an input (e.g., the local continuum flux [measured in the cloud's reference frame]) at a prior time. The first term in equation (3-3) is the component of $y$ that mirrors the variations in $x$ without delay, while the second term is the component of $y$ that responds on the time scale $\tau_y$.

For the important case in which the instantaneous component of the gain is zero, equations (3-2)-(3-3) yield the results indicated earlier in this section: in the fast cloud regime they yield an output that mirrors the input variations, while in the slow cloud regime they yield an output that is constant.

### 3.2. The Line Profile

Now that we have prescribed a general way of accounting for individual cloud properties that exhibit hysteresis-like behavior, we can derive the more observable properties of AGN models which have finite (rather than zero) equilibrium times. Of particular interest here is the angle-dependent apparent luminosity $L_l^{cl}$ emitted in line $l$ of a cloud with position vector from the black hole $\mathbf{r}$ and velocity vector $\mathbf{v}$. An expression for this that is general enough for the models that will be analyzed in this series of papers and which takes local delays into account is

$$L_l^{cl}(t, \mathbf{r}, \mathbf{v}, \hat{\mathbf{s}}) = F_c(t, \mathbf{r})A\epsilon_l[1 + \epsilon_{Al}\hat{\mathbf{r}} \cdot \hat{\mathbf{s}}] + L_l^s, \tag{3-4}$$

where $F_c(t, \mathbf{r})$ is the ionizing continuum flux at $\mathbf{r}$, $A$ is the cloud area, $\epsilon_l$ is the dimensionless emission efficiency for line $l$, $\epsilon_{Al}$ is the first-moment correction to the efficiency for an anisotropically-emitting cloud, $\mathbf{D} \equiv \mathbf{r} + \mathbf{s}$ is the position vector of the observer, and $L_l^{\mathbf{s}}$ is the cloud luminosity in line $l$ due to resonance scattering. Each of the cloud parameters in equation (3-4) that has an equilibrium time near or greater than $r/c$ (hereafter, the "spatial time" scale) must be evaluated using the appropriate form of equations (3-3) and (B6). Before a model conforming to equation (3-4) can have predictive power, not only must the continuum light curve be measured, but also estimates of the time scales and the asymptotic functional dependence of each cloud parameter upon the local continuum flux must be made.

Once a specific expression for the observed line flux from an individual cloud is assumed, the macroscopic characteristics of the global system composed of several clouds are easy to calculate. Neglecting absorption, the flux per cloud observable at $\mathbf{D}$ is $F_l^{\mathrm{cl}}(t, \mathbf{r}, \mathbf{v}; \mathbf{D}) \propto L_l^{\mathrm{cl}}(t - s/c, \mathbf{r}, \mathbf{v}, \hat{\mathbf{s}})/s^2$. With this terminology, the time-dependent line profile becomes (see Appendix A)

$$F_l(t, v_D) = \int d^3r \, d^3v \, f(\mathbf{r}, \mathbf{v}) F_l^{\mathrm{cl}}(t, \mathbf{r}, \mathbf{v}; \mathbf{D}) \delta(v_D + \mathbf{v} \cdot \hat{\mathbf{D}}), \qquad (3\text{-}5)$$

where $v_D$ is the equivalent tangential velocity and $f$ is the distribution function. Under the conditions specified earlier, the above equation permits computation of the line profile for any class of AGN cloud line emission models which can be described by equation (3-4).

Though equation (3-5) provides a means of computing the nonlinear line profile response when the input history is known, applying it can be computationally expensive (though not as much the trajectory-dependent sum method considered in Appendix A). This is because it requires modeling the cloud properties such as the $F_l^{\mathrm{cl}}$ function, which from a numerical perspective is an array with dimensions $\mathbf{r}$ and $\mathbf{v}$ that evolves with time, though probably only weakly in the $\mathbf{v}$ dimensions. If local delays are important, evaluation of $F_l^{\mathrm{cl}}$ at each point in time requires integrating over history according to the appropriate forms of equation (B6). Since the positional integral in the above equation can be interpreted as an integral over history, this makes the expression for the line profile a double integration over lag. This is in contrast with the analogous expression for the line profile given by equation (2.12) of Blandford & McKee (1982), which involves only a single integration over lag.

### 3.3. The Line Transfer Function and Linear Approximation of the Line Profile

By linearizing equation (3-4) (see Appendix B), one of the integrations in lag in equation (3-5) can be eliminated, and the computer time required to obtain the time-dependent line profile of a model can be significantly reduced. For several key cases of interest, we shall find that these benefits outweigh the inaccuracy of linear models.

The first step in linearizing the line flux emitted from an individual cloud is to obtain its gain about the bias continuum flux. This in turn requires determining the transfer function (eq. [3-2]) of each flux-dependent cloud parameter affecting the line emission in equation (3-4). Because the gain is calculated by considering small perturbations about the mean of the input, the gain of each of these parameters can be computed using the time-averaged local continuum flux, which is dependent only upon position. In terms of the gains of these cloud parameters, equation (3-4) yields for an individual cloud

$$\tilde{\Psi}_{L_l^{cl}|F_c}(\omega; \mathbf{r}, \mathbf{v}, \hat{\mathbf{s}}) = 1 + \tilde{\Psi}_{A|F_c}(\omega) + \tilde{\Psi}_{\epsilon_l|F_c}(\omega) + \frac{<\epsilon_{Al}> \hat{\mathbf{r}} \cdot \hat{\mathbf{s}}}{1 + <\epsilon_{Al}> \hat{\mathbf{r}} \cdot \hat{\mathbf{s}}} \tilde{\Psi}_{\epsilon_{Al}|F_c}(\omega) + \tilde{\Psi}_{L_l^s|F_c}(\omega), \quad (3\text{-}6)$$

where we will assume that the continuum flux is evaluated locally (at $\mathbf{r}$). Each of the above terms is proportional to the gain of one of the reactive cloud parameters. The third term itself is the sum of three highly model-dependent terms if $\epsilon_l = \epsilon_l(\Xi, P, N_c)$ sufficiently parameterizes the cloud emission for a given model. Note that even if the various cloud properties such as the area are described without approximation by a nontrivial linear response function, the output line response in a cloud is nonlinear nonetheless. This is because the above equation provides only an approximation to the response valid for small perturbations about a mean. Such nonlinearity is a general property of reactive cloud models.

With each cloud's response linearized, the remaining time dependence in the system line flux equation is due purely from the $F_c$ factor, so the global transfer function of the profile is

$$\tilde{\Psi}_{F_l|F_c}(\omega; v_D) = \int d^3r \, d^3v \frac{<F_l^{cl}(t, \mathbf{r}, \mathbf{v}; \mathbf{D})>_t}{<F_l>} f(\mathbf{r}, \mathbf{v}) \tilde{\Psi}_{F_l^{cl}|F_c}(\omega; \mathbf{r}, \mathbf{v}, \hat{\mathbf{s}}) \times$$
$$e^{-i\omega(\hat{\mathbf{r}} - \hat{\mathbf{D}}) \cdot \mathbf{r}/c} \delta(v_D + \mathbf{v} \cdot \hat{\mathbf{D}}). \quad (3\text{-}7)$$

The line flux gain of an individual cloud appearing in this equation is equal to the gain of the individual line luminosity (eq. [3-6]) if we neglect absorption. Note that unlike the Fourier transform of the expression given for the response function by Blandford & McKee (1982), the above equation has a factor of the cloud gain that can be nonzero at nonzero frequency. In the time domain, equation (3-7) gives (via eq. [B6]) the linear approximation to the line profile flux,

$$F_l(t, v_D) = <F_l(t, v_D)>_t \left[ 1 + \int d\tau \left( \frac{F_c(t - \tau, \mathbf{D})}{<F_c>} - 1 \right) \hat{\Psi}_{F_l|F_c}(\tau; v_D) \right] \pm \sigma_{F_l}, \quad (3\text{-}8)$$

where $\sigma_{F_l}$ represents the input-dependent error. Unlike equation (3-5), this equation does not have an implicit nested integration in lag. Applications of it to the models considered in this paper that account for the dependence of $F_l^{cl}$ upon the direction of the velocity vector are provided elsewhere (Paper IV; Taylor 1994).

It is worth repeating that a condition for the linearized response function to be descriptive is that at a given radius the continuum variations are small enough that the second order derivatives can be neglected. Since large-scale variations in the continuum luminosity are known to occur, the system would be somewhat contrived to consistently obey this condition. For instance, in the slow cloud regime the linearized form of equation (3-6) will generally be inaccurate at low enough mean cloud ionization parameters when the emitting ion is in partial fractional abundance. In this case, the second derivative of $\epsilon_l$ with respect to $F_c$ would not only be large and positive for most lines, but would also be a sensitive function of input level. If the asymptotic mean cloud ionization parameter is a decreasing function of the local flux ($s < 2$ [§ 1]), this implies overestimates for the size of the line emission region when using fully linear models. Even for wind cloud models (e.g., Kazanas 1989) in the fast cloud regime with $s = 2$, where the asymptotic value of the effective ionization parameter can be taken to be constant, nonlinearities would still arise from the dependence of the cloud area upon flux for which a constant $\eta(A|F_c)$ term cannot account. In any of these types of situations, the Fourier transform of the oscillatory component of the line flux would not be proportional to that of the continuum, and forms of equation (3-8) would not accurately describe the variability that would be observed.

In such cases, the "optimal" response function that best fits real data becomes a function of both the specific data set as well as the fitting criteria (see also eqs. [B12]-[B13]). Therefore, its utility in measuring any of the epoch-independent features of AGN and AGN models is somewhat questionable. This is in contrast with the linearized response function (eq. [3-7]), which is dependent upon only the mean of the continuum flux. Ideally, a fitting criteria would exist that would reliably yield this input-independent, poorly fitting linearized response function rather than the optimal one. However, there are alternative parameters and parameterized functions for analyses of variability data that completely bypass this problem. One alternative is the cross-correlation function. However, even in the linear regime this is also a strong function of the excitation characteristics. (See also § 4.) A more promising alternative is to fit nonlinear models (e.g., eq. [B8]; Taylor & Kazanas 1992) to data. Using nonlinear models offers the potential of epoch-independent fitting or measurement within the context of a model of physical AGN properties even when the continuum variations are large or the line emission is a sensitive function of the flux. This point will be discussed in more detail in Paper II.

## 4. EXAMPLES

Let us consider cloud models in which the effective reprocessing efficiencies are increasing functions of the local continuum flux, with the equilibrium time scale of the relevant physical properties being slightly larger than the characteristic light crossing time. From § 3, we know that responses to short and weak pulses of continuum radiation in such a system could be modeled satisfactorily with a linear "spatial" response function, which is the response function of the system were the fast cloud regime applicable. This response function has structure on just the light crossing times of the emission region, as the efficiency and physical conditions of the clouds in such a system would deviate only slightly from their mean values. Similarly, responses to long pulses of fixed intensity could also probably be mimicked with a *different* linear response function that had additional structure at lags beyond the cloud equilibrium times. However, if either short pulses and long pulses of constant intensity or long pulses of varying intensity occurred in such a system, a single linear response function would not be able to fit all aspects of the variability. A linear system would either respond too strongly to the weak pulses or too weakly to the strong pulses and furthermore would either respond too slowly to the weak pulses or too rapidly to the energetic pulses. The first two types of nonlinear behavior are due to an input-dependent asymptotic gain, while the latter two are due to nested lags or "inseparability" of the cloud and spatial response functions for systems in which the input is a multiplicative factor in the expression of the output.

These effects can be seen more clearly by considering a simple shell-like system in which the light crossing time is slightly shorter than the cloud equilibrium time. Specifically, let $f(\mathbf{r}, \mathbf{v}) \propto \delta(r - 10 \text{ days} \cdot c)$, the asymptotic cloud area function be $A'(F_c) = A_0(F_c/F_0)^\alpha$, $\tau_A = 30$ days, $\epsilon_{Al}=0$, $\epsilon_l = 1$, and the normalization of $f$ determined by the condition that the mean covering factor be unity, i.e., $< F_l >=< F_c >$. The ratio of the area equilibrium time to the light crossing time of this system is 3, which is near enough to 1 for neither the fast or slow cloud regime (§ 3) to be applicable. Exact outputs obtained upon application of equations (3-4)-(3-5) for various values of $\alpha$ of this system are displayed as solid lines $b$, $d$, and $e$ of Figure 1, while the input continuum that was assumed is shown as the solid line $a$. This input is a "low state" followed by "high state" that lasts for 200 days. Superimposed upon the low and high states are delta-function-like spikes of area 10 (luminosity-unit)-days, the responses of which can give a crude indication of a system's spatial response function. The outputs from the linearized response functions are shown as dotted lines. Ideally the optimal linearized response functions would have been obtained from a fitting scheme that minimized the discrepancy between the exact outputs. However, in this work they were obtained simply from equation (B4), equations (3-6)-(3-8), and finally equation (B13), which was derived for sinusoidal-like inputs but which results in surprisingly good fits for the input here as well.

For the nonreactive $\alpha = 0$ case shown in solid line $b$, the gain of the output line luminosity is unity, and the linearized response function of the system is just the spatial response function, which is a step function. Whether in the high or low state, here the amplitude of the response on time scales larger than the spatial time is the same as that of the input. The $\alpha = 1$ model is shown in solid line $d$. For this reactive model the cloud area responds linearly to the local continuum flux. This is shown explicitly in line $c$, which is the time-dependent area of the clouds on the shell after the spatial delay was removed by artificially setting $f(\mathbf{r}, \mathbf{v}) \propto \delta(r)$. Care must be taken in the interpretation of this response, as the actual size of this system is infinitely smaller than the size that the formalism of Blandford & McKee (1982) would yield, which is $\sim 40$ light-days$\sim c\tau_{\max}/2$, where $\Psi_{F_l|F_c}(\tau > \tau_{\max}) \simeq 0$. It is important to understand that although the cloud areas respond linearly in the $\alpha = 1$ model, the output itself is over-responsive from linear, with an asymptotic gain of 2 (eq. [3-6]). This asymptotic gain is also approximately the correction factor by which the formalism of Blandford & McKee (1982) would overestimate the cloud number density. (The exact factor is dependent upon the input, as eq. [B13] indicates, as well as the fitting criteria.) However, by taking into account an asymptotic gain correction factor that is different from unity, the linear response (dotted line $d$) does a surprisingly good job of fitting the actual response (solid line $d$) of the system, especially given that the input continuum luminosity function varies by a factor 7.5.

Nonlinear response is more apparent in the "under-responsive" $\alpha = -1$ model shown in solid line $e$. Here the cloud area response is given by a nonlinear input-modified system (eq. [B8]). A key difference between the models shown in solid lines $d$ and $e$ is that the gain of an individual cloud area is a decreasing function of positive frequency for the over-responsive model, but is an increasing function for the under-responsive model. Note that because the areas of the spikes are small, the actual responses to the first spike are similar in both cases, when the spatial response function (solid line $b$) does a crude job of describing the systems. However, low frequency or high (time-integrated) energy excitation exposes the latent nonlinearities of these systems. For instance, the over-responsive system (solid line $d$) responds slightly higher to the spike in the high state than the spike in the low state, while the linearized response function predicted a response that was the same strength for both spikes. This aspect of behavior is due to a nonlinear asymptotic gain. Even if it were accounted for, the shapes given by the linearized response function would still not perfectly match the exact ones. For instance, the linearized output for the over-responsive system responds too rapidly to the beginning of the high state. If its linearized response function were adjusted to yield a slower response, the linearized output would then respond too slowly to the beginning of the spike in the low state. Ultimately this is due to the area factor and hence the area equilibrium time playing a less important role in the response to the first spike of short duration (when the area is relatively constant) than in response to the energetic high state of long duration (when the area increases significantly). These types of problems are particularly evident for the highly nonlinear under-responsive case. Because the amplitude of the asymptotic gain of the area is only 1, the response to both low-energy spikes is square-like. However, the high state is energetic and long enough to permit the areas to respond and the asymptotic gain of zero nearly to be attained, which results in triangle-like responses. The linearized response function incorrectly gives triangle-like features in the responses to the spikes.

For the models shown in Figure 1, all the cloud property gain terms in equation (3-6) except that of the area are zero. However, for calculations of *relative* line strengths, the ionization parameter gain term in equation (3-6) frequently determines the key distinguishing response characteristics. This is illustrated by the models shown in Figure 2, where the response as a function of the pressure equilibrium time scale is shown for three models nearly identical to those used in Figure 1. However, for these cases, $P \propto F_c^{s/2}$, the cloud areas and column densities are forced to be constant, and the spike strengths are reduced. For model $a$, the pressure equilibrium time is 1.1 days, which is approximately the mean inverse Strömgren sound crossing time for these clouds. This is short enough (compared to the spatial time of 10 days) for the fast cloud regime to be valid. Thus, in this case, a single (rather than double) integration in lag would have sufficed for calculating $L_l(t)$. For model $b$, the pressure equilibrium time is 500 days, which is closer to the thermal evaporation time scale (Table 1) of $\sim 2 \times 10^3$ days. In the limit that the pressure equilibrium time scale becomes infinite, model $b$ is identical to a nonreactive $s = 0$ model, which also would not require the computationally expensive double integration. Finally, for model $c$ the equilibrium time scale is 30 days. This intermediate equilibrium time might be applicable for cloud models in which the evaporation rate is a strong function of the mean ionization state (see also Taylor 1994). Because the column density to pressure ionization parameter ratio ($1.2 \times 10^{22}$ cm$^{-2}$) was selected to place the model near the "C IV—limited" state in which the instantaneous component of the C IV gain is $\sim -1$, the C IV line responds very weakly to the initial spike. However, it responds strongly on the pressure equilibrium time scale to the beginning of the extended high state, during which time the pressure regulation mechanism readjusts the ionization parameter partially back toward its lower initial equilibrium value. In contrast, at the ending of the high state, the C IV line drops on the shorter spatial time scale because of the higher pressures (and lower ionization parameters) of the clouds. This example illustrates one of several ways in which the mean response time of a system can be dependent upon the recent mean continuum luminosity. Response features like these will be discussed in much more detail in Paper III (Taylor, in preparation 1996) within the context of more complex models.

The above examples clearly illustrate how the global linearized response function can have structure not due to the spatial response function. However, nonlinear effects can even mask the spatial information. For example, if the ratio of the local continuum flux to the product of the mean pressure and column density of the clouds becomes high enough, the emission of certain lines could be "recombination-limited," which makes $\hat{\tilde{\Psi}}_{c_l|F_c}(\infty) \sim -1$ in equation (3-6). This occurs during the high state with Ly$\alpha$ for model $b$ (solid line Fig. 2). Therefore, as the model shown in the solid line $c$ of Figure 1 also illustrates, the response times give, under the assumption of spherical symmetry, only *upper* limits to the characteristic size.

The above examples also help illustrate how local delays can affect the cross-correlation function. Consider a simple $\eta(A|F_c) > 0$ model, such as that shown in solid line $d$ of Figure 1. For a weak enough high-frequency square wave input occurring above a steady input background, the output would be a capped triangle wave, which is symmetric about its peak with respect to lag. The cross-correlation function that one would obtain from such data would be shifted of order 10 days, yet also be symmetric about the characteristic lag. This is because the data from the beginning, or growing phase of the pulse, which alone would produce a cross-correlation function with a positive slope, is compensated by the ending or falling phase of the response pulse.

Consider, however, the case where the input is a moderate-intensity, low-frequency square wave. Because of local delays, the response would be higher near the end of the pulse, as in solid line $d$ of Figure 1. Since the cross-correlation function is an amplitude-biased function, the resulting cross-correlation function would be biased from the data in the falling phase. Unlike the previous case, this would result in a negatively-sloped component to the cross-correlation function, which is quite common (e.g., Sparke 1993).

Note that local delays only give the cross correlation function asymmetry about their peak above what one would obtain from the global linearized response function alone. This is because the cross-correlation function is the convolution of the response function and the symmetric, input auto-correlation function. For this reason, assessing the importance of asymmetric cross-correlation functions in determining the characteristics of the local delays of interest here would probably require knowledge of the best-fitting linear response function.

Note that this function would be biased somewhat by the low-energy (more symmetric) pulses. Therefore, the asymmetry of the "simulated cross-correlation function" that one would obtain from the output of this response function would not be as great as the one obtained from actual variability data. Thus, the difference between the simulated and actual cross-correlation functions can be asymmetric if local delays are important. This permits a simple way for testing for local delays. More complicated techniques will be discussed in other papers in this series.

## 5.   SUMMARY AND CONCLUSIONS

In this paper it has been argued that for some reactive cloud models the fast cloud assumption is invalid. In such cases a tight correspondence between the positional distribution of matter and the global linearized line flux response function simply does not exist. This is because the continuum luminosity at a given time in history affects not only the clouds on a shell, but also the clouds inside the sphere such a shell bounds. In general, the response which results is not only nonlinear, but also inseparable, requiring more than one integration over lag to determine the output flux at a specified time. However, in some cases the time-dependence of the line fluxes can be described using linearized response functions $\Psi_{F_l|F_c}$. These response functions have structure at intermediate lags due to the finite size of the line-emitting region and at lags greater than these due to the finite equilibrium times of the line-emitting material itself. For small enough perturbations the physical cloud properties can be relatively static and a linear response function can work quite well at describing the responses. However, when the input continuum variations become extreme enough, such response functions can fail. Because of nonlinear asymptotic response, the integral of an observable's linearized response function differs from the time average of the observable by a correction factor of the asymptotic gain. Ignoring nonlinear effects can lead to incorrect measurements of the physical properties of the system, such as sizes that are too large.

One of the most fundamental of assumptions that has been made in previous analysis of variability data has been the fast cloud assumption. With this assumption now in question, this data should be examined again with models that do not require it.

It is my pleasure to thank L. Titarchuk, D. Kazanas, T. Kallman, A. Raval, B. Piner, and H. Netzer for offering suggestions that improved the presentation of ideas in this paper. This work was supported by NASA Cooperative Agreement NCC 5-54 and is part of a dissertation to be submitted to the Graduate School, University of Maryland in partial fulfillment of the requirements of the Ph. D. degree in Physics.

## A.  The Line Flux of a Cloud with an Equilibrium Time of Order of the Crossing Time

In standard works such as Blandford & McKee (1982), the time-dependent line profile is an integral over phase space of the distribution function and the flux emitted per cloud. The cloud fluxes themselves are functions of the various cloud properties, which are assumed to be only a function of position and the instantaneous local continuum flux. As hinted in § 3.1, if the equilibrium times are of order of the cloud crossing times (which is apparently the case for most of the cloud models that do not require extensive cloud creation throughout the emission region), the cloud properties have an additional explicit dependence upon the prior continuum fluxes $F_c(t - \tau, \mathbf{r}(t - \tau))$ and hence the orbital trajectories of the clouds, which is a different function for each cloud. This would complicate modeling efforts, which would entail integrating over the orbital trajectories of the clouds, and would make the much simpler approach taken by Blandford in McKee (1982) invalid. However, in this section we shall find that in certain cases the cloud luminosity function can be described merely by giving the properties of the cloud a velocity dependence in addition to the intrinsic ones.

Let us first obtain the exact solution to this problem. Ignoring absorption, the (nonlinear) line flux as a function of time for a cloud is given by application of equation (3-3) to each of the cloud parameters in equation (3-4). Thus the continuum-subtracted, time-dependent line profile of the "global" system observed from $\mathbf{D}$ is

$$F_i(t, v_D) = \sum_{i=1}^{N} F_{li}^{\text{orb}}(t) \delta(v_D + \mathbf{v}_i \cdot \hat{\mathbf{D}}), \tag{A1}$$

where $N$ is the number of clouds and $F_{li}^{\text{orb}}$ is the flux in line $l$ from cloud $i$ computed from the forms of equations (3-3)-(3-4) appropriate for the model to be tested. Neglecting absorption and non-Doppler line broadening, equation (A1) gives the exact time-dependent line profile for clouds with arbitrary motions. However, because the suspected number of clouds in AGN is high (e.g., Laor et al. 1994; c.f., e.g., Peterson 1994), using it could prove computationally expensive.

Noting the exponential factor in equation (3-3), integration over the history of local continuum exposure required for determining the cloud properties at given time need only be carried out to a small factor (e.g., $\sim 4$) of the relevant equilibrium time. Therefore, if each cloud property relevant to emission has an equilibrium time scale that is appreciably less than the emission region crossing time, the radius will not change drastically over the relevant history interval, and the continuum flux function can approximated by its Taylor expansion. This yields a first order correction to the continuum flux function that is proportional to the radial velocity. Similar expansions permit estimation of the line flux observed from an individual cloud and the line shifts that equation (A1) implies, as will be described in detail in Paper IV.

However, if $N$ is independent of the continuum flux and is large enough that line broadening produces a smooth line profile, a more accurate method for obtaining the individual cloud line flux that partially accounts for higher order terms can be obtained simply by taking the statistical average the function, which is

$$F_l^{\text{cl}}(t, \mathbf{r}, \mathbf{v}; \mathbf{D}) = \lim_{\delta r, \delta v \to 0; N \to \infty} \frac{1}{f(\mathbf{r}, \mathbf{v}) \delta^3 r \delta^3 v} \times$$

$$\sum_{i=1}^{N} \int_{\mathbf{r}, \mathbf{v}}^{\mathbf{r} + \delta \mathbf{r}, \mathbf{v} + \delta \mathbf{v}} d^3 r' d^3 v' \delta(\mathbf{r}' - \mathbf{r}_i(t)) \delta(\mathbf{v}' - \mathbf{v}_i(t)) F_{li}^{\text{orb}}(t). \tag{A2}$$

The dependence of $F_l^{\text{cl}}(t, \mathbf{r}, \mathbf{v}; \mathbf{D})$ upon time is due to the variation of the continuum flux to which the cloud is locally exposed. The dependence upon position is due to traditional model elements such as changes in the mean cloud density as a function of average heating. Finally, the dependence upon the velocity vector accounts for the intrinsic dependence as well as that due to the history of heating being important when the equilibrium time scale is not completely negligible compared to the emission region crossing time. Note that for models where the position and velocity variables impose the integrals of motion of a cloud's trajectory, the above condition that the equilibrium times are small compared to the crossing times is unnecessary and the flux is an exact function of only the time, position, and velocity variables.

In this case, an analog of equation (A2) can be used to replace the knowledge of the individual cloud trajectories with the time-independent phase space distribution function $f(\mathbf{r}, \mathbf{v})$. Though this function, when combined with equation (A2), permits equation (3-5) to be used to obtain an approximation of the time-dependent line profile, it offers little advantage over using just equation (A1) because it still entails explicit time-dependent orbital modeling. However, in the linear regime (eqs. [3-7]-[3-8]), only the time-average of equation (A2) is required to obtain the time-dependent line profile. Once this (velocity-dependence) has been computed for a model, the linear approximations to the observable characteristics can be obtained from equation (3-8) for various continuum light curves without explicit time-dependent orbital modeling.

## B.    Response of Nonlinear Systems in the Linear Regime

In Blandford & McKee (1982), linear systems were analyzed in the linear regime. In this appendix a formalism is developed for analyzing *non*linear systems in the linear regime. Though the solution to this problem is a straightforward and probably necessary prerequisite for any comprehensive understanding of variability in AGN, it was not correctly obtained or applied in other works regarding AGN variability. We shall find that within the linear regime the analysis in Blandford & McKee (1982) is inadequate for general nonlinear systems.

Let us consider the generic system described in § 3.1. Using the notation of § 3.1 for $y(t), y'(x), x(t)$, and $t$, there is no reason that $\delta y'(x) \propto \delta x$ should generally hold, and one may be forced to employ a fully nonlinear analysis method to accurately describe the system. However, let us assume here that the variations in $x$ are sufficiently smaller than its mean, in which case, provided $y'(x)$ is a smooth function, it can be approximated with

$$y' = <y> \left[ 1 + \eta(y|x) \left( \frac{x}{<x>} - 1 \right) \right] \pm \sigma_y, \qquad (B1)$$

where the dimensionless, "asymptotic gain" $\eta$ of changes in $y$ for small and slow changes in $x$ is defined by

$$\eta(y|x) \equiv \lim_{\delta x, \dot{x} \to 0} \frac{\delta y/y}{\delta x/x} = \frac{<x>}{<y>} \frac{\partial y'}{\partial x}|_{x=<x>}, \qquad (B2)$$

where $\sigma_y$ is the error due to nonzero second order derivatives in $y'$. The asymptotic gain has elsewhere been termed the "responsivity" (Krolik et al. 1991; Goad, O'Brien, & Gondhalekar 1993). Here it is an operator to distinguish between the various gains with the different "output" and "input" functions that will be required, though note that it is independent of the normalizations of these functions.

Let us extend the definition of gain by allowing a dependence upon the type of input signal. Consider a time-dependent local ionizing continuum flux or a Fourier component of it such as $x(t) - x_0 = x_1 \cos(\omega t)$. A dimensionless frequency-dependent gain or transfer function $\tilde{\tilde{\Psi}}$ of $y(t)$ with respect to $x(t)$ can then be defined as

$$\tilde{\tilde{\Psi}}_{y|x}(\omega) \equiv \lim_{\delta x \to 0} \frac{\delta y/y}{\delta x/x}|_{\ddot{x}/x_1 = -\omega^2}. \qquad (B3)$$

Letting $\tilde{y}(\omega)$ denote the Fourier transform of $y(t)$, etc., this gives

$$\tilde{\tilde{\Psi}}_{y|x}(\omega) = \frac{<x>}{<y>} \frac{\partial \tilde{y}(\omega)}{\partial \tilde{x}(\omega)} \qquad (B4)$$

and an analog of the Fourier transform of equation (B1) for a frequency-dependent gain,

$$\tilde{y}(\omega) \simeq <y> \left[ \delta(\omega) + \tilde{\tilde{\Psi}}_{y|x}(\omega) \left( \frac{\tilde{x}(\omega)}{<x>} - \delta(\omega) \right) \right]. \qquad (B5)$$

With this notation, $|\hat{\tilde{\Psi}}_{y|x}(\omega)|$ is the dimensionless ratio of the amplitudes of variations of $y$ to $x$, while $-\mathrm{Im}[\ln{(\hat{\tilde{\Psi}}_{y|x}(\omega))}]/\omega$ is the delay in response.[2] Similarly, the asymptotic gain is $\mathrm{Re}[\hat{\tilde{\Psi}}_{y|x}(0)]$, while the "instantaneous component of the gain" is $\mathrm{Re}[\hat{\tilde{\Psi}}_{y|x}(\infty)]$.

Equation (B5) yields in the time domain

$$y(t) = <y> \left[ 1 + \int d\tau \left( \frac{x(t-\tau)}{<x>} - 1 \right) \hat{\Psi}_{y|x}(\tau) \right] \pm \sigma_y, \tag{B6}$$

where we define the inverse Fourier transform of the gain of the output $y(t)$ with respect to the input $x(t)$ as the normalized linearized response function, which is

$$\Psi_{y|x}(\tau) \equiv \frac{<y>}{<x>} \hat{\Psi}_{y|x}(\tau). \tag{B7}$$

Here the lack of a caret denotes that the normalized response function has scaling other than that given to it by the inverse Fourier transform. Note that the integral of the linearized response function is just the asymptotic gain, which is only unity for actual linear systems.

Linearization is not always advantageous, as in some cases, including those obeying equation (3-1), the exact solution can easily be obtained from

$$y(t) = \int d\tau \, y'(x(t-\tau)) \Psi_{y|y'}(\tau), \tag{B8}$$

where

$$\Psi_{y|y'}(\tau) = \hat{\Psi}_{y|y'}(\tau) = \hat{\Psi}_{y|x}(\tau)/\eta(y|x) \tag{B9}$$

is the response function of an "input-modified" system.

However, linearization can be quite useful, as in some cases it allows complex systems to be accurately described by a single equivalent response function, which can drastically reduce the simulation time. Such is the case where one is interested in obtaining an observable quantity of a system with many clouds, where the convolution of $y$ with a spatial linear response function must be evaluated. For instance, consider a hypothetical system where a physical cloud property $z'(y) \propto y^{\eta(z|y)}$ lags property $y$ on a time scale $\omega_z^{-1}$, while property $y'(x) \propto x^{\eta(y|x)}$ has a "direct" lag of $\omega_y^{-1}$. The exact expression for $z(t)$ has two nested integrals over lags. However, upon linearization the transfer function of $z$ for small variations in $x$ is

$$\hat{\tilde{\Psi}}_{z|x}(\omega) = \frac{\eta(z|y)}{1 + i\omega/\omega_z} \frac{\eta(y|x)}{1 + i\omega/\omega_y}, \tag{B10}$$

or alternatively

$$\hat{\Psi}_{z|x}(\tau) = \Theta(\tau)\eta(z|y)\eta(y|x)\omega_z\omega_y \left( \frac{e^{-\omega_z\tau}}{\omega_y - \omega_z} + \frac{e^{-\omega_y\tau}}{\omega_z - \omega_y} \right) \qquad \{\omega_y \neq \omega_z\}$$

$$= \Theta(\tau)\eta(z|y)\eta(y|x)\omega_z^2\tau e^{-\omega_z\tau} \qquad\qquad \{\omega_y = \omega_z\}, \tag{B11}$$

which when applied (in eq. [B6]) requires only a single integral over lag. For future reference, note that when $\omega_z \gg \omega_y$ or $\omega_z \ll \omega_y$, the two gain factors are "separable" from one another, i.e. for a restricted range of excitation frequencies one of the gain factors can be treated as a constant.

---

[2]The sign in the Fourier transform used here, though different from that in several references (e.g., Blandford & McKee 1982), minimizes differences with the Laplace transform, which offers certain advantages in dealing with this type of problem.

In this section it has been shown (eq. [B9]) that there is a "correction factor" of $\eta(y|x)$ in the expression for the "gain-corrected response function." Previous works (e.g., Blandford & McKee 1982) assumed that the systems themselves are linear, which is equivalent to assuming correction factors of unity. Some of the problems with making this assumption are pointed out in Goad, O'Brien, & Gondhalekar (1993) as well as § 4 of this work. Note that the correction factors differ from unity in equation (B2) in nonlinear systems even if the perturbations are arbitrarily small and equation (B6) accurately describes the system.

Partially accounting for even higher order corrections due to nonlinearity is also possible within the linear regime and, in fact, is important for accurate interpretation of fits of linear models to nonlinear systems. This is because if the variations are not infinitesimal, the above equations do not necessarily yield the "optimal" fit that would have obtained using real variability data. For instance, consider the case where $y' = y_0(x/x_0)^\alpha$. The better-fitting optimal average for a sinusoidal-like input is the first Fourier coefficient of $y'$,

$$<y> = \frac{1}{2\pi} \int_0^{2\pi} d(\omega t) y'(x_0 + x_1\cos(\omega t)) \neq y'(<x>). \tag{B12}$$

Similarly, the observable asymptotic gain is approximately

$$\eta(y|x) = \frac{1}{\pi} \int_0^{2\pi} d(\omega t)\cos(\omega t)\frac{y'(x_0 + x_1\cos(\omega t))}{<y>} \neq \alpha, \tag{B13}$$

where the inequalities can be removed only for the $x_1 \ll x_0$ case.

TABLE 1

EQUILIBRIUM PROCESSES OF SOME REACTIVE CLOUD
MODEL PARAMETERS AND THEIR EFFECTS

| Name of Limiting Process | Cloud Parameters Affected | Parameter Equilibrium Times | Observational Parameters Affected |
|---|---|---|---|
| Inter-cloud Cooling[1] | $A, N_c, P_l$ | 43 days | RF, LR (strong) |
| Thermal Evaporation[2] | $A, N_c, P_l$ | 20 years | RF, LP, LR (strong) |
| Pressure-limited Evaporation[3] | $A, N_c$ | 2.0 days | RF, LR (weak) |
| Pressure-limited Evaporation[3,4] | $P_l$ | $\sim 0.37$ days | RF, LR (strong) |
| Stellar Wind Expansion[5] | $A, N_c, P_l$ | 20 days | RF, LR (strong) |
| Stellar Photospheric Heating[6] | $A, N_c, P_l$ | $10^{-2} - 10^{11}$ days | RF, LP, LR (strong) |
| Magnetic Confinement[7] | $A, N_c, P_l$ | 4.3 days | RF, LR (strong) |

NOTE—"RF," "LP," and "LR" are respective abbreviations for "response functions," "line profiles," and "line ratios." These results are for clouds at a fiducial radius from the continuum source $r_0$ of 10 light-days (the light crossing time scale), a fiducial local continuum flux of $10^{44}/(4\pi r_0^2)$ ergs cm$^{-2}$, a fiducial velocity of 4000 km s$^{-1}$, a fiducial cloud hydrogen density of $10^{11}$ cm$^{-3}$, and a fiducial mean column density of $10^{22.5}$ cm$^{-2}$.

[1]Only for models with clouds in pressure equilibrium with a hot inter-cloud medium (e.g., Krolik, McKee, & Tarter 1981). Calculation assumes an inter-cloud temperature of $10^7$ K, which implies that the dominant source of cooling is thermal Bremsstrahlung, which in turn implies flux-dependent (reactive) cloud parameters.

[2]Adopted from results in Krinsky and Puetter (1992), but after scaling to the column density assumed here. Line ratios are only strongly affected for pressure-stratified clouds.

[3]Only for pressure-stratified cloud models, see Taylor (1994).

[4]Only for lines emitted uniformly from the inverse-Strömgren region.

[5]Adopted from parameters assumed in Schaaf & Schmutzler (1992).

[6]Adapted from Harpaz & Rappaport (1991) and Antona & Ergma (1993).

[7]As in Rees (1987), but assuming the field responds to the continuum flux on the Alvén wave cloud crossing time.

Fig. 1.—Comparison of the linear approximation to the actual responses of simple shell-like systems with local delays. Line $a$ is the input continuum that was assumed, which has a luminosity of 0.5 in arbitrary units, the "low state," followed by a luminosity of 1.25 units, the "high state," which lasts for 200 days. Superimposed upon the low and high states are delta-function-like spikes of area 10 unit-days. The solid lines $b - e$ are the output line luminosities for the models described in the text (§ 4) offset respectively by -1, -1.75, -3, and -5 luminosity units while the dotted lines are approximations of the outputs obtained from linearized response functions. Though the linearized responses do a reasonable job of matching the actual responses for most of the models shown here, they fail to exhibit the differences between weak and strong (time-integrated) excitation. This is particularly evident for the model shown in solid line $e$.

Fig. 2.—Effect of the pressure equilibrium time upon the "line-specific" response. The top line (left axis) is the input continuum luminosity assumed. The exact Ly$\alpha$ (solid lines) and C IV (dotted lines) output luminosities (left axis) for also shown for three extremely simple $s = 1$ models similar to those shown in Fig. 1. To emphasize the effect of just the cloud pressures and pressure ionization parameters being locally delayed, the cloud areas were artificially forced to yield a constant geometrical covering factor of unity (neglecting absorption) and the cloud column densities were artificially forced constant at $10^{22}$ cm$^{-2}$. The initial column density to pressure ionization parameter ratios assumed were $1.2 \times 10^{22}$ cm$^{-2}$. The line luminosities for models $b$ and $c$ are offset respectively by $-10^{42}$ and $-3 \times 10^{42}$ ergs s$^{-1}$. The pressure-equilibrium time $\tau_{P_l}$ assumed in models $a, b$, and $c$ was respectively 1.1, 500, and 30 days. For pressure-stratified clouds, this respectively corresponds to wave propagation speeds of $\sim c_s$ (pressure-limited evaporation), $\sim 2 \times 10^{-3} c_s$ (thermal evaporation), and $\sim 4 \times 10^{-2} c_s$ (intermediate evaporation), where $c_s$ is the sound speed. The photoionization code that was used is XSTAR (see, e.g., Kallman 1995). The spectrum that was assumed is shown in Fig. 3.

Fig. 3.—The spectrum that was assumed for the models shown in Fig. 2. It is identical to that used in Krolik et al. (1991). The $y$-axis has been plotted in linear (as opposed to logarithmic) coordinates.

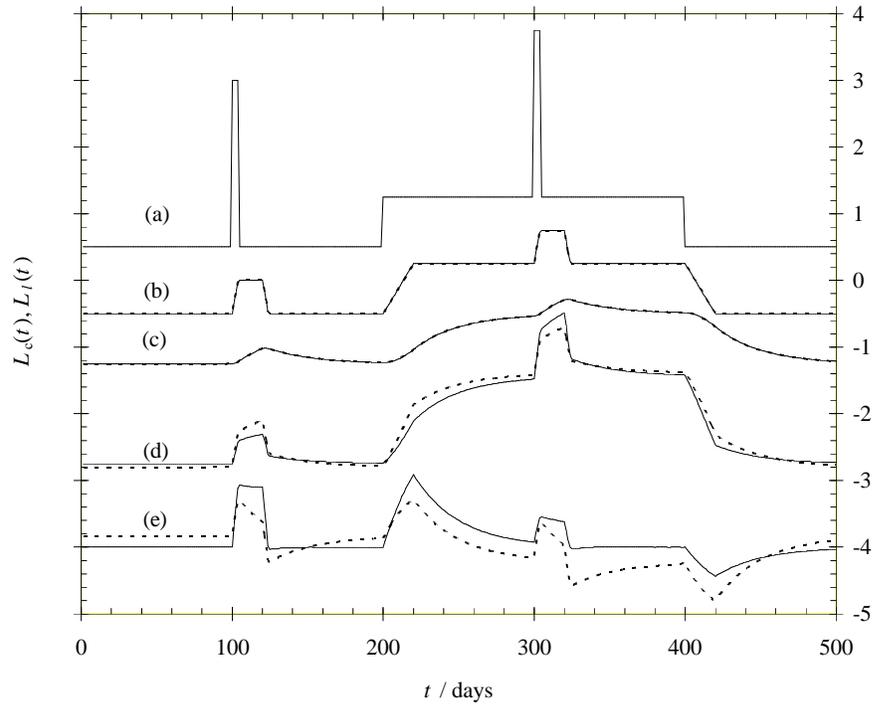

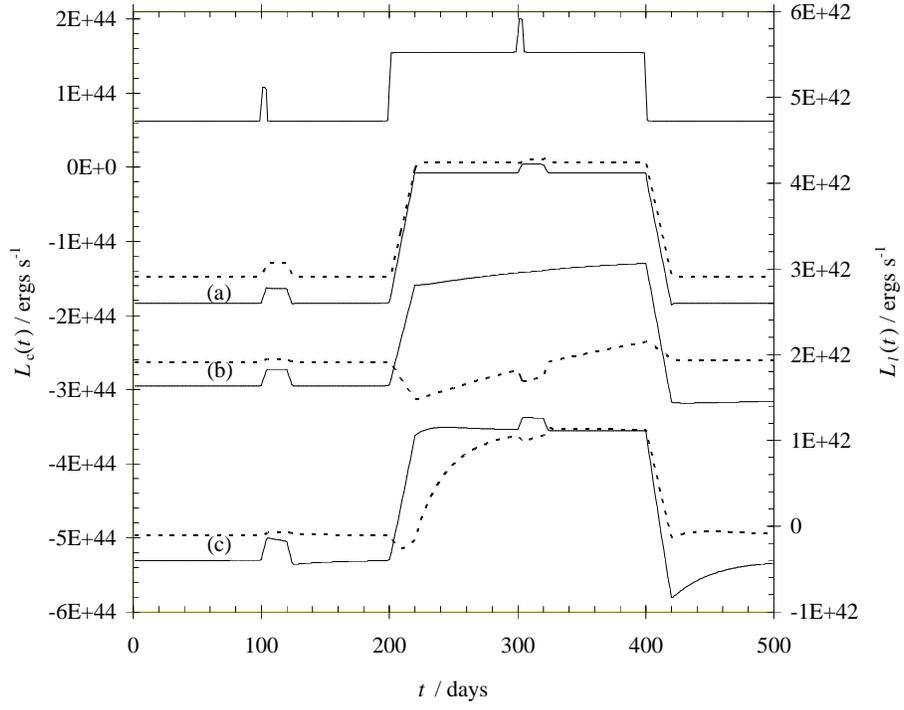

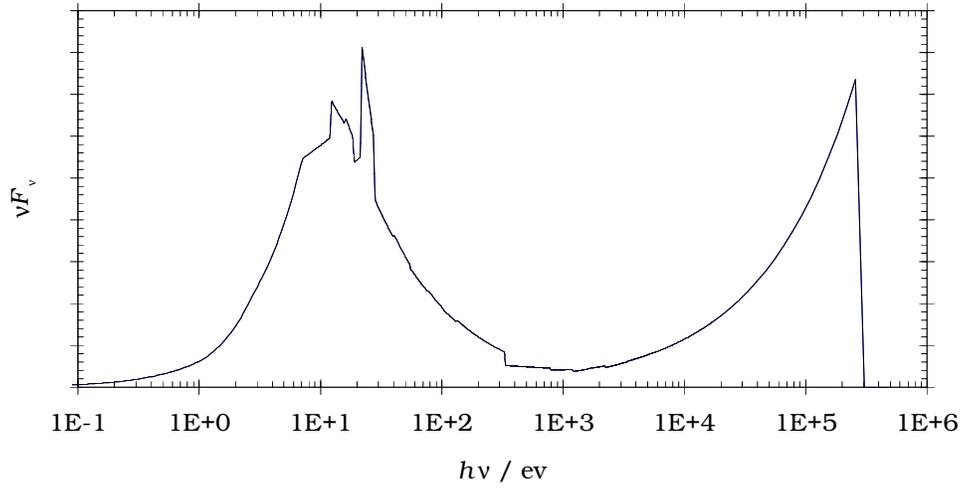